\documentclass[onecolumn]{revtex4}
\usepackage{amsfonts}
\usepackage{amsmath}

\begin{document}

\title{Unifying the theory of Integration within normal-, Weyl- and
antinormal-ordering of operators and the s--ordered operator
expansion formula of density operators }
\author{Hong-yi Fan \\{\small Department of Physics, Shanghai Jiao Tong University, Shanghai 200030,
China}\\fhym@sjtu.edu.cn.}

\begin{abstract}
By introducing the $s$--parameterized generalized Wigner operator
into phase-space quantum mechanics we invent the technique of
integration within $s$--ordered product of operators (which
considers normal ordered, antinormally ordered and Weyl ordered
product of operators as its special cases). The $s$--ordered
operator expansion (denoted by $\S \cdots \S)$ formula of density
operators is derived, which is
\[
\rho=\frac{2}{1-s}\int \frac{d^{2}\beta}{\pi}\left \langle -\beta
\right \vert \rho \left \vert \beta \right \rangle \S \exp \{
\frac{2}{s-1}\left( s|\beta|^{2}-\beta^{\ast}a+\beta
a^{\dagger}-a^{\dagger}a\right)  \} \S,
\]
The $s$--parameterized quantization scheme is thus completely established.


\textbf{PACC:} 0365, 0530, 4250

\end{abstract}

\maketitle

\section{Introduction}

The subject about operators and their classical correspondence has
been a hot topic since the birth of quantum mechanics (QM) and now
becomes a field named QM in phase space. Because Heisenberg's
uncertainty principle prohibits the notion of a system being
described by a point in phase space, only domains of minimum area
$2\pi \hbar$ in phase space is allowed. Wigner \cite{r1} introduced
a function whose marginal distribution gives probability of a
particle in coordinate space or in momentum space, respectively. The
Wigner distribution is related to operators' Weyl ordering (or Weyl
quantization scheme) \cite{r2}. We notice that each phase space
distribution is associated with a definite operator ordering for
quantizing classical functions. For examples, P-representation (as a
density operator $\rho$'s classical correspondence) is actually
$\rho$'s antinormally ordered expansion in terms of the completeness
of coherent state $\left \vert z\right \rangle =\exp
[-\frac{|z|^{2}}{2}+za^{\dagger}]\left \vert 0\right \rangle $
\cite{r3,r4},
\begin{equation}
\rho=\int \frac{d^{2}z}{\pi}P\left(  z\right)  \left \vert z\right \rangle
\left \langle z\right \vert \label{1}%
\end{equation}
because the coherent states compose a complete set $\int \frac{d^{2}z}{\pi
}\left \vert z\right \rangle \left \langle z\right \vert =1$ \cite{r5}. The Wigner
distribution function $W\left(  p,x\right)  $ of $\rho,$ defined as $Tr\left[
\rho \Delta \left(  p,x\right)  \right]  ,$ is proportional to the classical
Weyl correspondence $h\left(  p,x\right)  $ of $\rho$ ($\rho$'s Weyl ordered
expansion), i.e.,%
\begin{equation}
\rho=\iint \limits_{-\infty}^{\infty}dpdx\Delta \left(  p,x\right)  h\left(
p,x\right)  ,\label{2}%
\end{equation}%
\begin{equation}
Tr\left[  \rho \Delta \left(  p,x\right)  \right]  =\left(  2\pi \right)
^{-1}h\left(  p,x\right)  =W\left(  p,x\right)  .\label{3}%
\end{equation}
since the Wigner operator $\Delta \left(  p,x\right)  $ is complete too,
$\iint \limits_{-\infty}^{\infty}dpdx\Delta \left(  p,x\right)  =1.$ The
original form of $\Delta \left(  p,x\right)  $ defined in the coordinate
representation is \cite{r6}
\begin{equation}
\Delta \left(  x,p\right)  =\int \limits_{-\infty}^{\infty}\frac{du}{2\pi
}e^{iup}\left \vert x+\frac{u}{2}\right \rangle \left \langle x-\frac{u}%
{2}\right \vert ,\label{4}%
\end{equation}
for the Wigner operator in the entangled state representation we
refer to
\cite{r7}. When $\rho=\left(  \frac{1}{2}\right)  ^{m}\sum \limits_{l=0}%
^{m}\binom{m}{l}X^{m-l}P^{n}X^{l}$, $\left[  X,P\right]  =i$,
$\hbar=1$, according to Eqs. (\ref{3})-(\ref{4}), the classical
correspondence of $\left(
\frac{1}{2}\right)  ^{m}\sum \limits_{l=0}^{m}\binom{m}{l}X^{m-l}P^{n}X^{l}$ is%
\begin{align}
& 2\pi Tr\left[  \left(  \frac{1}{2}\right)  ^{m}\sum \limits_{l=0}^{m}%
\binom{m}{l}X^{m-l}P^{n}X^{l}\Delta \left(  x,p\right)  \right]  \nonumber \\
& =\int_{-\infty}^{\infty}due^{ipu}\left \langle x-\frac{u}{2}\right \vert
\left(  \frac{1}{2}\right)  ^{m}\sum \limits_{l=0}^{m}\frac{m!}{l!\left(
m-l\right)  !}X^{m-l}P^{r}X^{l}\left \vert x+\frac{u}{2}\right \rangle
\nonumber \\
& =x^{m}\int_{-\infty}^{\infty}due^{ipu}\left \langle x-\frac{u}{2}\right \vert
P^{r}\left \vert x+\frac{u}{2}\right \rangle \nonumber \\
& =x^{m}\int_{-\infty}^{\infty}due^{ipu}\int_{-\infty}^{\infty}dp^{\prime
}e^{-ip^{\prime}u}p^{\prime r}\nonumber \\
& =x^{m}\int_{-\infty}^{\infty}dp^{\prime}\delta \left(  p-p^{\prime}\right)
p^{\prime r}\nonumber \\
& =x^{m}p^{r},\label{5}%
\end{align}
this is the original definition of Weyl quantization scheme
(quantizing classical coordinate and momentum quantity $x^{m}p^{n}$
as the corresponding operators) as \cite{r2}
\begin{equation}
x^{m}p^{n}\rightarrow \left(  \frac{1}{2}\right)  ^{m}\sum_{l=0}^{m}\binom
{m}{l}X^{m-l}P^{n}X^{l},\label{6}%
\end{equation}
its right-hand side is in Weyl ordering, so we introduce the symbol $%
%TCIMACRO{\QATOP{:}{:}}%
%BeginExpansion
\genfrac{}{}{0pt}{}{:}{:}%
%EndExpansion%
%TCIMACRO{\QATOP{:}{:}}%
%BeginExpansion
\genfrac{}{}{0pt}{}{:}{:}%
%EndExpansion
$ to characterize it \cite{r8}, i.e.,%
\begin{equation}
\left(  \frac{1}{2}\right)  ^{m}\sum_{l=0}^{m}\binom{m}{l}X^{m-l}P^{n}X^{l}=%
%TCIMACRO{\QATOP{:}{:}}%
%BeginExpansion
\genfrac{}{}{0pt}{}{:}{:}%
%EndExpansion
\left(  \frac{1}{2}\right)  ^{m}\sum_{l=0}^{m}\binom{m}{l}X^{m-l}P^{n}X^{l}%
%TCIMACRO{\QATOP{:}{:}}%
%BeginExpansion
\genfrac{}{}{0pt}{}{:}{:}%
%EndExpansion
,\label{7}%
\end{equation}
It is worth emphasizing that the order of operators $X$ and $P$ are permuted
within the Weyl ordering symbol \cite{r8}, a useful property which has been
overlooked for a long time, Based on this fact a useful method called
integration within Weyl ordered product of operators has been invented
\cite{r8}.

Therefore, from Eq. (\ref{6}) and Eq. (\ref{7})
\begin{equation}
x^{m}p^{r}\rightarrow%
%TCIMACRO{\QATOP{:}{:}}%
%BeginExpansion
\genfrac{}{}{0pt}{}{:}{:}%
%EndExpansion
\left(  \frac{1}{2}\right)  ^{m}\sum_{l=0}^{m}\binom{m}{l}X^{m-l}P^{n}X^{l}%
%TCIMACRO{\QATOP{:}{:}}%
%BeginExpansion
\genfrac{}{}{0pt}{}{:}{:}%
%EndExpansion
=%
%TCIMACRO{\QATOP{:}{:}}%
%BeginExpansion
\genfrac{}{}{0pt}{}{:}{:}%
%EndExpansion
X^{m}P^{r}%
%TCIMACRO{\QATOP{:}{:}}%
%BeginExpansion
\genfrac{}{}{0pt}{}{:}{:}%
%EndExpansion
.\label{8}%
\end{equation}
Following Eq. (\ref{11}) we have
\begin{equation}%
%TCIMACRO{\QATOP{:}{:}}%
%BeginExpansion
\genfrac{}{}{0pt}{}{:}{:}%
%EndExpansion
X^{m}P^{r}%
%TCIMACRO{\QATOP{:}{:}}%
%BeginExpansion
\genfrac{}{}{0pt}{}{:}{:}%
%EndExpansion
=\int \int_{-\infty}^{\infty}dpdx\Delta \left(  x,p\right)  x^{m}p^{r},\label{9}%
\end{equation}
which implies $\Delta \left(  x,p\right)  =%
%TCIMACRO{\QATOP{:}{:}}%
%BeginExpansion
\genfrac{}{}{0pt}{}{:}{:}%
%EndExpansion
\delta \left(  x-X\right)  \delta \left(  p-P\right)
%TCIMACRO{\QATOP{:}{:}}%
%BeginExpansion
\genfrac{}{}{0pt}{}{:}{:}%
%EndExpansion
$, or $\Delta \left(  \alpha \right)  =\frac{1}{2}%
%TCIMACRO{\QATOP{:}{:}}%
%BeginExpansion
\genfrac{}{}{0pt}{}{:}{:}%
%EndExpansion
\delta \left(  \alpha^{\ast}-a^{\dagger}\right)  \delta \left(  \alpha-a\right)
%
%TCIMACRO{\QATOP{:}{:}}%
%BeginExpansion
\genfrac{}{}{0pt}{}{:}{:}%
%EndExpansion
$, $\alpha=\left(  x+ip\right)  /\sqrt{2},$ a delta operator-function form in
Weyl ordering.

Having realized that each phase space distribution accompanies a definite
operator ordering for quantizing classical functions, we may think of that
each complete set of operators corresponds to an operator-ordering rule. In
this work we shall introduce a complete set of operators characteristic of a
$s$-parameter (the generalized Wigner operator) and then introduce a
generalized quantization scheme with the $s$-parameter operator ordering.
Historically, Cahill and Glauber \cite{r9} have introduced the $s$%
--parameterized quasiprobability distribution according to which the
coherent state expectation of $\rho$, the Wigner function of $\rho,$
and the P-representation of $\rho$ respectively corresponds to three
distinct values of $s,$ i.e., $s=1,0,-1.$ However, the
$s$--parameterized quantization scheme associated with the
$s$--parameterized quasiprobability distribution has not been
completely established, as the fundamental problem of what is
$\rho$'s $s$-ordered operator expansion has not been touched yet. In
another word, the problem of how to arrange any given operator as
its $s$-ordered form has been unsolved, say for instance, no
references has ever reported what is the $s$-ordered operator
expansion of $exp\left(  \lambda a^{\dagger}a\right)  ?$ ($\left[
a,a\right]  =1$) In this work we shall solve this important problem
by introducing the technique of integration within $s$-ordering of
operators, which in the cases of $s=1,0,-1,$ respectively goes to
the technique of integration within normal-ordering, Weyl ordering
and antinormal ordering of operators. In this way we can tackle
these three techniques in a unified way. The work is arranged as
follows: In Sec. 2 we introduce the explicit $s$--parameterized
Wigner operator $\Delta_{s}\left(  \alpha \right)  $ and then in
Sec. 3 we establish one-to-one mapping between operators and their
$s$--parameterized classical correspondence after proving the
relation $2\pi Tr\left[  \Delta_{-s}\left(  \alpha^{\prime
\ast},\alpha^{\prime}\right) \Delta_{s}\left(  \alpha^{\ast},\alpha
\right)  \right]  =\delta \left( x^{\prime}-x\right)  \left(
p^{\prime}-p\right)  $, where $\alpha=\left( x+ip\right) /\sqrt{2}.$
In Sec. 4 we introduce the symbol $\S \cdots \S$ denoting
$s$--ordering of operators and the technique of integration within
$s$--ordered product of operators. In Sec. 5-6 we derive density
operator's expansion formula in terms of $s$--ordered quantization
scheme, such that the $s$-ordered expansion of $exp\left( \lambda
a^{\dagger }a\right)  $ is obtained. In this way we develop and
enrich the theory of phase space quantum mechanics.

\section{The s--parameterized Wigner operator and quantization scheme}

Our aim is to construct s--parameterized quantization scheme, in another word,
we want to construct a one-to-one correspondence between an operator and its
classical correspondence in the sense of s--parameterized quasiprobability
distribution. For this purpose we should introduce a generalized Wigner
operator for the s--parameterized phase space theory. By analogy with the
usual Wigner operator \cite{r6} we introduce a generalized Wigner operator for
$s$--parameterized distributions,
\begin{equation}
\Delta_{s}\left(  \alpha \right)  =\int \frac{d^{2}\beta}{2\pi^{2}}\exp \left(
\frac{s|\beta|^{2}}{2}+\beta a^{\dagger}-\beta^{\ast}a-\beta \alpha^{\ast
}+\beta^{\ast}\alpha \right)  .\label{10}%
\end{equation}
Using the Baker-Hausdorff formula to put the exponential in normally ordered
form, and using the technique of integration within normal product of
operators \cite{r10,r11}, for $s<1,$ we obtain%
\begin{align}
\Delta_{s}\left(  \alpha \right)    & =\int \frac{d^{2}\beta}{2\pi^{2}}%
\colon \exp \left[  \frac{-\left(  1-s\right)  |\beta|^{2}}{2}+\beta a^{\dagger
}-\beta^{\ast}a-\beta \alpha^{\ast}+\beta^{\ast}\alpha \right]  \colon
\nonumber \\
& =\frac{1}{\left(  1-s\right)  \pi}\colon \exp \left[  \frac{-2}{1-s}\left(
a^{\dagger}-\alpha^{\ast}\right)  \left(  a-\alpha \right)  \right]
\colon,\label{11}%
\end{align}
this is named s--parameterized Wigner operator. In particular, when $s=0,$ Eq.
(\ref{11}) reduces to the usual normally ordered Wigner operator \cite{r12}%
\begin{align}
\Delta_{s}\left(  \alpha \right)    & \rightarrow \Delta \left(  \alpha \right)
=\frac{1}{\pi}\colon \exp \left[  -2\left(  a^{\dagger}-\alpha^{\ast}\right)
\left(  a-\alpha \right)  \right]  \colon \nonumber \\
& =\frac{1}{\pi}\colon \exp \left[  -\left(  x-X\right)  ^{2}-\left(
p-P\right)  ^{2}\right]  \colon,\label{12}%
\end{align}
where $X=\frac{a^{\dagger}+a}{\sqrt{2}},P=\frac{i\left(  a^{\dagger}-a\right)
}{\sqrt{2}}.$ On the other hand, by putting the exponential in (\ref{10})
within antinormal ordering symbol $\vdots$ $\vdots,$ we have for $s<-1,$%
\begin{align}
\Delta_{s}\left(  \alpha \right)    & =\int \frac{d^{2}\beta}{2\pi^{2}}%
\vdots \exp \left[  \frac{-\left(  -1-s\right)  |\beta|^{2}}{2}+\beta
a^{\dagger}-\beta^{\ast}a-\beta \alpha^{\ast}+\beta^{\ast}\alpha \right]
\vdots \nonumber \\
& =\frac{1}{\left(  -1-s\right)  \pi}\vdots \exp \left[  \frac{2}{1+s}\left(
a^{\dagger}-\alpha^{\ast}\right)  \left(  a-\alpha \right)  \right]
\vdots.\label{13}%
\end{align}
In reference to the asymptotic expression of Delta function $\delta \left(
x\right)  =\lim_{\epsilon \rightarrow0}\frac{1}{\sqrt{\pi \epsilon}}%
e^{-\frac{x^{2}}{\epsilon}},$ we have for $s=-1,$%
\begin{align}
\Delta_{s=-1}\left(  \alpha \right)    & =\vdots \delta \left(  a^{\dagger
}-\alpha^{\ast}\right)  \delta \left(  a-\alpha \right)  \vdots=\delta \left(
a-\alpha \right)  \delta \left(  a^{\dagger}-\alpha^{\ast}\right)  \nonumber \\
& =\left \vert \alpha \right \rangle \left \langle \alpha \right \vert ,\label{14}%
\end{align}
which is the pure coherent state, $\left \vert \alpha \right \rangle =\exp \left(
-\frac{1}{2}|\alpha|^{2}+\alpha a^{\dagger}\right)  \left \vert 0\right \rangle
.$ Using
\begin{equation}
\colon \exp \left[  \left(  e^{\lambda}-1\right)  a^{\dagger}a)\right]
\colon=e^{\lambda a^{\dagger}a},\label{15}%
\end{equation}
we can convert (\ref{11}) to the form%
\begin{align}
\Delta_{s}\left(  \alpha \right)    & =\frac{1}{\left(  1-s\right)  \pi
}e^{\frac{2}{1-s}\alpha a^{\dagger}}\colon \exp \left[  \left(  \frac{s+1}%
{s-1}-1\right)  a^{\dagger}a\right]  \colon e^{\frac{2}{1-s}\alpha^{\ast
}a-\frac{2}{1-s}|\alpha|^{2}}\nonumber \\
& =\frac{1}{\left(  1-s\right)  \pi}e^{\frac{2}{1-s}\alpha a^{\dagger}%
}e^{a^{\dagger}a\ln \frac{s+1}{s-1}}e^{\frac{2}{1-s}\alpha^{\ast}a-\frac
{2}{1-s}|\alpha|^{2}}.\label{16}%
\end{align}

\section{The $s$--parameterized quantization scheme}

It follows from (\ref{11}) that%
\begin{equation}
2\int d^{2}\alpha \Delta_{s}\left(  \alpha \right)  =\frac{2}{\left(
1-s\right)  \pi}\int d^{2}\alpha \colon \exp \left[  \frac{-2}{1-s}\left(
a^{\dagger}-\alpha^{\ast}\right)  \left(  a-\alpha \right)  \right]
\colon=1,\label{17}%
\end{equation}
so $\Delta_{s}\left(  \alpha \right)  $ is complete and $\rho$ can be expanded
as
\begin{equation}
\rho=2\int d^{2}\alpha \Delta_{s}\left(  \alpha \right)  \mathfrak{P}\left(
\alpha \right)  ,\label{18}%
\end{equation}
which is a new classic-quantum mechanical correspondence between
$\mathfrak{P}\left(  \alpha \right)  $ and $\rho$, when $s=0$, (\ref{18})
yields the Weyl correspondence. By noting the form of $\Delta_{-s}\left(
\alpha^{\prime}\right)  $ and using $\int \frac{d^{2}z}{\pi}\left \vert
z\right \rangle \left \langle z\right \vert =1$ we calculate%
\begin{align}
Tr\left[  \Delta_{-s}\left(  \alpha^{\prime}\right)  \Delta_{s}\left(
\alpha \right)  \right]    & =G\int \frac{d^{2}z}{\pi}\left \langle z\right \vert
e^{\frac{2}{1+s}\alpha^{\prime}a^{\dagger}}e^{a^{\dagger}a\ln \frac{s-1}{s+1}%
}e^{\frac{2}{1+s}\alpha^{\prime \ast}a}e^{\frac{2}{1-s}\alpha a^{\dagger}%
}e^{a^{\dagger}a\ln \frac{s+1}{s-1}}e^{\frac{2}{1-s}\alpha^{\ast}a}\left \vert
z\right \rangle \nonumber \\
& =G\int \frac{d^{2}z}{\pi}\left \langle z\right \vert e^{\frac{2}{1+s}%
\alpha^{\prime}a^{\dagger}}e^{\frac{-2}{1-s}\alpha^{\prime \ast}a}e^{\frac
{-2}{1+s}\alpha a^{\dagger}}e^{\frac{2}{1-s}\alpha^{\ast}a}\left \vert
z\right \rangle \nonumber \\
& =Ge^{\frac{4\alpha^{\prime \ast}\alpha}{\left(  1+s\right)  \left(
1-s\right)  }}\int \frac{d^{2}z}{\pi}\exp[\frac{2z^{\ast}}{1+s}\left(
\alpha^{\prime}-\alpha \right)  -\frac{2z}{1-s}\left(  \alpha^{\prime \ast
}-\alpha^{\ast}\right)  ]\nonumber \\
& =\frac{1}{4\pi}\delta \left(  \alpha^{\prime}-\alpha \right)  \left(
\alpha^{\prime \ast}-\alpha^{\ast}\right)  e^{-\left(  \frac{2}{1-s}+\frac
{2}{1+s}\right)  |\alpha|^{2}-\frac{4\alpha^{\prime \ast}\alpha}{\left(
1+s\right)  \left(  s-1\right)  }}\nonumber \\
& =\frac{1}{4\pi}\delta \left(  \alpha^{\prime}-\alpha \right)  \left(
\alpha^{\prime \ast}-\alpha^{\ast}\right)  \nonumber \\
& =\frac{1}{2\pi}\delta \left(  q^{\prime}-q\right)  \left(  p^{\prime
}-p\right)  ,\label{19}%
\end{align}
where $G\equiv \frac{e^{-\frac{2}{1-s}|\alpha|^{2}-\frac{2}{1+s}|\alpha
^{\prime}|^{2}}}{\left(  1+s\right)  \left(  1-s\right)  \pi^{2}}.$ Therefore,
the classical function corresponding to $\rho$ (in the context of the
$s$--parameterized quantization scheme) is given by%
\begin{align}
2\pi Tr\left[  \Delta_{-s}\left(  \alpha \right)  \rho \right]    & =4\pi \int
d^{2}\alpha^{\prime}Tr\left[  \Delta_{-s}\left(  \alpha \right)  \Delta
_{s}\left(  \alpha^{\prime}\right)  \right]  \mathfrak{P}\left(
\alpha^{\prime},s\right)  \nonumber \\
& =\int d^{2}\alpha^{\prime}\delta \left(  \alpha-\alpha^{\prime}\right)
\left(  \alpha^{\ast}-\alpha^{\prime \ast}\right)  \mathfrak{P}\left(
\alpha^{\prime},s\right)  \nonumber \\
& =\mathfrak{P}\left(  \alpha,s\right)  .\label{20}%
\end{align}

Eq. (\ref{20}) is the reciprocal relation of (\ref{18}). Thus we have
established one-to-one mapping between operators and their $s$%
--parameterized classical correspondence. The $s$--parameterized
quantization scheme is completed, of which the Weyl quantization is
its special case.

\section{Expansion formula of $\left \vert z\right \rangle \left \langle
z\right \vert $ in terms of $s$--parameterized quantization scheme}

When $\rho=\left \vert z\right \rangle \left \langle z\right \vert ,$ using
(\ref{20}) we have%
\begin{align}
2\pi Tr\left[  \Delta_{-s}\left(  \alpha \right)  \left \vert z\right \rangle
\left \langle z\right \vert \right]    & =\frac{2}{1+s}\left \langle z\right \vert
\colon \exp \left[  \frac{-2}{1+s}\left(  a^{\dagger}-\alpha^{\ast}\right)
\left(  a-\alpha \right)  \right]  \colon \left \vert z\right \rangle \nonumber \\
& =\frac{2}{1+s}\exp \left[  \frac{-2}{1+s}\left(  z^{\ast}-\alpha^{\ast
}\right)  \left(  z-\alpha \right)  \right]  ,\label{21}%
\end{align}
this is the $s$-parameterized classical correspondence of $\left \vert
z\right \rangle \left \langle z\right \vert $ in phase space. Eq. (\ref{21})
represents a kind of phase space distribution$,$ since the integration over it
leads to the completeness
\begin{equation}
\int \frac{d^{2}z}{\pi}\left \vert z\right \rangle \left \langle z\right \vert
\rightarrow \frac{2}{1+s}\int \frac{d^{2}z}{\pi}\exp \left[  \frac{-2}%
{1+s}\left(  z^{\ast}-\alpha^{\ast}\right)  \left(  z-\alpha \right)  \right]
=1.\label{22-1}%
\end{equation}
For this $s$-parameterized distribution we can define $s$-ordered form of
$\left \vert z\right \rangle \left \langle z\right \vert $ through the following
formula%
\begin{equation}
\left \vert z\right \rangle \left \langle z\right \vert =\frac{2}{1+s}%
\S \exp \left[  \frac{-2}{1+s}\left( z^{\ast}-a^{\dagger}\right)
\left(  z-a\right)  \right]  \S,\label{23}%
\end{equation}
where $\S \cdots \S$ means $s$-ordering symbol. This definition is
consistent with those well-known ordered formulas of $\left \vert
z\right \rangle \left \langle z\right \vert $. Indeed, when in
(\ref{23}) $s=0,$
$\S \cdots \S$ converts to Weyl ordering $%
%TCIMACRO{\QATOP{:}{:}}%
%BeginExpansion
\genfrac{}{}{0pt}{}{:}{:}%
%EndExpansion
\cdots%
%TCIMACRO{\QATOP{:}{:}}%
%BeginExpansion
\genfrac{}{}{0pt}{}{:}{:}%
%EndExpansion
$, so (\ref{23}) reduces to%
\begin{equation}
\left \vert z\right \rangle \left \langle z\right \vert =2%
%TCIMACRO{\QATOP{:}{:}}%
%BeginExpansion
\genfrac{}{}{0pt}{}{:}{:}%
%EndExpansion
\exp \left[  -2\left(  z^{\ast}-a^{\dagger}\right)  \left(  z-a\right)
\right]
%TCIMACRO{\QATOP{:}{:}}%
%BeginExpansion
\genfrac{}{}{0pt}{}{:}{:}%
%EndExpansion
,\label{24}%
\end{equation}
as expected \cite{r7}; when in (\ref{23}) $s=1,$ $\S \cdots \S$
becomes normal ordering \cite{r10},%
\begin{equation}
\left \vert z\right \rangle \left \langle z\right \vert =\colon \exp \left[
-\left(  z^{\ast}-a^{\dagger}\right)  \left(  z-a\right)  \right]
\colon,\label{25}%
\end{equation}
which is as expected too; when $s=-1,$ $\S \cdots \S$ becomes
antinormal ordering,%
\begin{align}
\left \vert z\right \rangle \left \langle z\right \vert  &  =\lim_{\epsilon
\rightarrow0}\frac{2}{\epsilon}\vdots \exp \left[  \frac{2}{\epsilon}\left(
z^{\ast}-a^{\dagger}\right)  \left(  z-a\right)  \right]  \vdots \nonumber \\
&  =\vdots \delta \left(  z^{\ast}-a^{\dagger}\right)  \delta \left(  z-a\right)
\vdots,
\end{align}
still is as expected.

\section{The technique of integration within $s$--ordered product of
operators}

Let us introduce the technique of integration within $s$-ordered product of
operators (IWSOP) by listing some properties of the $s$-ordered product of
operators which is defined through (\ref{23}):

1. The order of Boson operators $a$ and $a^{\dagger}$ within a
$s$-ordered symbol can be permuted, even though $\left[
a,a^{\dagger}\right]  =1.$

2. $c$-numbers can be taken out of the symbol $\S \cdots \S \ $as
one wishes.

3. An $s$-ordered product of operators can be integrated or differentiated
with respect to a $c$-number provided the integration is convergent.

4. The vacuum projection operator $|0\rangle \langle0|$ has the $s$-ordered
product form (see (\ref{23}))
\begin{equation}
|0\rangle \langle0|=\frac{2}{1+s}\S \exp \left(
\frac{-2}{1+s}a^{\dagger
}a\right)  \S. \label{27}%
\end{equation}

5. the symbol $\S \cdots \S$ becomes $:$ $:$ for $s=1$, becomes $%
%TCIMACRO{\QATOP{:}{:}}%
%BeginExpansion
\genfrac{}{}{0pt}{}{:}{:}%
%EndExpansion%
%TCIMACRO{\QATOP{:}{:}}%
%BeginExpansion
\genfrac{}{}{0pt}{}{:}{:}%
%EndExpansion
$ for $s=0$, and becomes $\vdots \  \vdots$ for $s=-1.$

\section{Density operator's expansion formula in terms of $s$--ordered
quantization scheme}

Using (\ref{1}) and (\ref{23}) we have the expansion within $\S
\cdots \S,$
\begin{equation}
\rho=\int \frac{d^{2}z}{\pi}P\left(  z\right)  \left \vert z\right
\rangle \left \langle z\right \vert =\frac{2}{1+s}\int
\frac{d^{2}z}{\pi}P\left( z\right)  \S \exp \left[
\frac{-2}{1+s}\left(  z^{\ast}-a^{\dagger
}\right)  \left(  z-a\right)  \right]  \S.\label{28}%
\end{equation}
Substituting Mehta's expression of $P\left(  z\right)  $ \cite{r13}
\begin{equation}
P\left(  z\right)  =e^{|z|^{2}}\int \frac{d^{2}\beta}{\pi}\left \langle
-\beta \right \vert \rho \left \vert \beta \right \rangle e^{|\beta|^{2}+\beta
^{\ast}z-\beta z^{\ast}},\label{29}%
\end{equation}
where $\left \vert \beta \right \rangle $ is also a coherent state, $\left \langle
-\beta \right \vert \left.  \beta \right \rangle =e^{-2|\beta|^{2}}$, into
(\ref{28}) we have%
\begin{eqnarray}
\rho &  =\frac{2}{1+s}\int \frac{d^{2}\beta}{\pi}\left \langle
-\beta \right \vert \rho \left \vert \beta \right \rangle
e^{|\beta|^{2}}\int \frac{d^{2}z}{\pi
}\S \exp \left[  |z|^{2}+\beta^{\ast}z-\beta z^{\ast}-\frac{2}%
{1+s}\left(  z^{\ast}-a^{\dagger}\right)  \left(  z-a\right)  \right]
\S \nonumber \\
&  =\frac{2}{1-s}\int \frac{d^{2}\beta}{\pi}\left \langle -\beta
\right \vert \rho \left \vert \beta \right \rangle \S \exp \left[
\frac{2}{s-1}\left( s|\beta|^{2}-\beta^{\ast}a+\beta
a^{\dagger}-a^{\dagger}a\right)  \right]
\S,\label{30}%
\end{eqnarray}
this is density operator's expansion formula in terms of
$s$--ordered
quantization scheme. In particular, when $s=0,$ (\ref{30}) becomes%
\begin{equation}
\rho=2\int \frac{d^{2}\beta}{\pi}\left \langle -\beta \right \vert \rho \left \vert
\beta \right \rangle
%TCIMACRO{\QATOP{:}{:}}%
%BeginExpansion
\genfrac{}{}{0pt}{}{:}{:}%
%EndExpansion
\exp \left[  2\left(  \beta^{\ast}a-\beta a^{\dagger}+a^{\dagger}a\right)
\right]
%TCIMACRO{\QATOP{:}{:}}%
%BeginExpansion
\genfrac{}{}{0pt}{}{:}{:}%
%EndExpansion
,\label{31}%
\end{equation}
which is the formula converting $\rho$ into its Weyl ordered form
\cite{r7,r8}; while for $s=-1,$ (\ref{30}) becomes%
\begin{equation}
\rho=2\int \frac{d^{2}\beta}{\pi}\left \langle -\beta \right \vert \rho \left \vert
\beta \right \rangle \vdots \exp \left[  -\left(  |\beta|^{2}+\beta^{\ast}a-\beta
a^{\dagger}+a^{\dagger}a\right)  \right]  \vdots,\label{32}%
\end{equation}
which is the formula converting $\rho$ into its antinormally ordered form
\cite{r14}, as expected.

\section{Application}

We now use (\ref{30}) to derive the $s$--ordered expansion of $e^{\lambda
a^{\dagger}a},$ using (\ref{15}) and the IWSOP technique we have%
\begin{eqnarray}
e^{\lambda a^{\dagger}a}  & =\frac{2}{1-s}\int \frac{d^{2}\beta}{\pi
}\left \langle -\beta \right \vert \exp \left[  \left(
1-e^{\lambda}\right)
|\beta|^{2}\right]  \left \vert \beta \right \rangle \S \exp \{ \frac{2}%
{s-1}\left(  s|\beta|^{2}-\beta^{\ast}a+\beta a^{\dagger}-a^{\dagger}a\right)
\} \S \nonumber \\
& =\frac{2}{1-s}\int \frac{d^{2}\beta}{\pi}\S \exp \left[ \left(
-1-e^{\lambda}\right)  |\beta|^{2}+\frac{2}{s-1}\left(  s|\beta|^{2}%
-\beta^{\ast}a+\beta a^{\dagger}-a^{\dagger}a\right)  \right] \S
\nonumber \\
& =\frac{2}{1+s-se^{\lambda}+e^{\lambda}}\S \exp \left[
\frac{2\left( e^{\lambda}-1\right)
}{1+s-se^{\lambda}+e^{\lambda}}a^{\dagger}a\right]
\S,\label{33}%
\end{eqnarray}

which is a new formula. For $s=1,$ $\S \cdots \S \rightarrow:$ $:$,
(\ref{33}) reduces to (\ref{15}) as expected; for $s=0,$ $\S
\cdots \S \rightarrow%
%TCIMACRO{\QATOP{:}{:}}%
%BeginExpansion
\genfrac{}{}{0pt}{}{:}{:}%
%EndExpansion%
%TCIMACRO{\QATOP{:}{:}}%
%BeginExpansion
\genfrac{}{}{0pt}{}{:}{:}%
%EndExpansion
$, (\ref{33}) becomes the Weyl ordering expansion \cite{r7}$,$%
\begin{equation}
e^{\lambda a^{\dagger}a}=\frac{2}{1+e^{\lambda}}%
%TCIMACRO{\QATOP{:}{:}}%
%BeginExpansion
\genfrac{}{}{0pt}{}{:}{:}%
%EndExpansion
\exp \left[  \frac{2e^{\lambda}-2}{1+e^{\lambda}}aa^{\dagger}\right]
%TCIMACRO{\QATOP{:}{:}}%
%BeginExpansion
\genfrac{}{}{0pt}{}{:}{:}%
%EndExpansion
,\label{34}%
\end{equation}
and for $s=-1,$ $\S \cdots \S \rightarrow \vdots \ \vdots,$
(\ref{33}) becomes\cite{r10}%
\begin{equation}
e^{\lambda a^{\dagger}a}=e^{-\lambda}\vdots \exp \left[  \left(  1-e^{-\lambda
}\right)  aa^{\dagger}\right]  \vdots,
\end{equation}
which is also correct. Further, we consider the $s$--ordered expansion of the
generalized Wigner operator itself, using (\ref{10}) and (\ref{30}) we have%
\begin{eqnarray}
\Delta_{s}\left(  \alpha^{\ast},\alpha \right)    & =\frac{2}{\left(
1-s\right)  ^{2}\pi}\int \frac{d^{2}\beta}{\pi}\left \langle -\beta
\right \vert \colon \exp \left[  \frac{-2}{1-s}\left(
a^{\dagger}-\alpha^{\ast}\right) \left(  a-\alpha \right)  \right]
\colon \left \vert \beta \right \rangle
\nonumber \\
& \times \S \exp \left[  \frac{2}{s-1}\left(
s|\beta|^{2}-\beta^{\ast
}a+\beta a^{\dagger}-a^{\dagger}a\right)  \right]  \S \nonumber \\
& =\frac{2}{\left(  1-s\right)  ^{2}\pi}\int
\frac{d^{2}\beta}{\pi}\S e^{-2|\beta|^{2}}\exp \left \{
\frac{2}{s-1}\left[  \left(  -\beta^{\ast }-\alpha^{\ast}\right)
\left(  \beta-\alpha \right)  \right.  \right.
\nonumber \\
& \left.  \left.  +s|\beta|^{2}-\beta^{\ast}a+\beta a^{\dagger}-a^{\dagger
}a\right]  \right \}  \S \nonumber \\
& =\frac{2}{\left(  1-s\right)  ^{2}}\S \delta \left[  \frac{2}%
{s-1}\left(  a^{\dagger}-\alpha^{\ast}\right)  \right]  \delta \left[  \frac
{2}{s-1}\left(  a-\alpha \right)  \right]  \S,\label{35}%
\end{eqnarray}
which in the case of $s=0$ becomes the Weyl ordered form of the
usual Wigner
operator $\Delta \left(  \alpha \right)  =\frac{1}{2}%
%TCIMACRO{\QATOP{:}{:}}%
%BeginExpansion
\genfrac{}{}{0pt}{}{:}{:}%
%EndExpansion
\delta \left(  \alpha^{\ast}-a^{\dagger}\right)  \delta \left(  \alpha-a\right)
%
%TCIMACRO{\QATOP{:}{:}}%
%BeginExpansion
\genfrac{}{}{0pt}{}{:}{:}%
%EndExpansion
.$

In summary, by introducing the $s$--parameterized generalized Wigner
operator into phase-space quantum mechanics we have proposed the
technique of integration within $s$--ordered product of operators
(which considers normal ordered, antinormally ordered and Weyl
ordered product of operators as its special cases). The $s$--ordered
operator expansion (denoted by $\S \cdots \S$) formula of density
operators is derived. The theory of Integration within normal-,
Weyl- and antinormal-ordering of operators can now be tackled in a
unified way. The $s$--parameterized quantization scheme is
completely established, of which the Weyl quantization is its
special case. For the mutual transformation between the Weyl
ordering and $X-P$ (or $P-X$) ordering of operators we refer to
\cite{r15}.

\end{document}